\documentclass[a4paper,12pt]{article}
\usepackage[sumlimits,intlimits,namelimits]{amsmath} 
\usepackage[english]{babel}
\usepackage{graphicx}
\usepackage[pdftex]{color}
\usepackage{geometry}
\makeatletter
\def\fmslash{\@ifnextchar[{\fmsl@sh}{\fmsl@sh[0mu]}}
\def\fmsl@sh[#1]#2{%
  \mathchoice
    {\@fmsl@sh\displaystyle{#1}{#2}}%
    {\@fmsl@sh\textstyle{#1}{#2}}%
    {\@fmsl@sh\scriptstyle{#1}{#2}}%
    {\@fmsl@sh\scriptscriptstyle{#1}{#2}}}
\def\@fmsl@sh#1#2#3{\m@th\ooalign{$\hfil#1\mkern#2/\hfil$\crcr$#1#3$}}
\makeatother
\numberwithin{equation}{section}
\newcommand{\beq}{\begin{equation}}
\newcommand{\eeq}{\end{equation}}
\newcommand{\bea}{\begin{eqnarray}}
\newcommand{\eea}{\end{eqnarray}}

\newcommand{\tto}{\!\to\!}
\newcommand{\msp}[1]{\mbox{\hspace*{#1mm}~}}

\newcommand{\GeV}{\,\mbox{GeV}}

\newcommand{\matel}[3]{\langle #1|#2|#3\rangle}
\newcommand{\aver}[1]{\langle #1\rangle}

\newcommand{\Lam}{\Lambda_{\rm QCD}}

\newcommand{\gsim}{\lower.7ex\hbox{$
\;\stackrel{\textstyle>}{\sim}\;$}}
\newcommand{\lsim}{\lower.7ex\hbox{$
\;\stackrel{\textstyle<}{\sim}\;$}}

\newcommand{\bibit}[1]{\bibitem{#1}}

\begin{document}
\begin{titlepage}
\setcounter{page}{0}
\thispagestyle{empty}
\begin{flushright}
SI-HEP-2009-14 \\
UND-HEP-09-BIG03 \\[0.2cm]
\end{flushright}

\vspace{1.2cm}
\begin{center}
{\Large\bf 
The Two Roads to `Intrinsic Charm' in $B$ Decays}
\end{center}

\vspace{0.5cm}
\begin{center}
{\sc Ikaros Bigi $^a$}, 
{\sc Thomas Mannel $^b$, Sascha Turczyk $^b$, Nikolai Uraltsev $^{a,b,c}$} \\[0.1cm]
{\sf  $^a$ Dept.\ of Physics, University of Notre Dame du Lac, Notre Dame, 
IN 46556, U.S.A.} \\[0.1cm] 
{\sf $^b$ Theoretische Physik 1, Fachbereich Physik,
Universit\"at Siegen\\ D-57068 Siegen, Germany}\\[0.1cm] 
{\sf  $^c$ St.\,Petersburg Nuclear Physics Institute, Gatchina,
St.\,Petersburg 188300, Russia}
\end{center}

\vspace{0.8cm}
\begin{abstract}
\vspace{0.2cm}\noindent
We describe two complementary ways to show the presence of higher order
effects in the $1/m_Q$ expansion for inclusive $B$ decays that have been
dubbed `Intrinsic Charm'. Apart from the lessons they can teach us about QCD's
nonperturbative dynamics their consideration is relevant for precise
extractions of $|V_{cb}|$: for they complement the estimate of the potential
impact of $1/m_Q^4$ contributions. We draw semiquantitative conclusions for
the expected scale of Weak Annihilation in semileptonic $B$ decays, both for
its valence and non-valence components.
\end{abstract}
\vfill

\tableofcontents

\end{titlepage}

\newpage

\section{Introduction}

The Operator Product Expansion (OPE) has turned out to be a highly powerful
tool for describing inclusive decays of beauty hadrons. Among the major
results and successes are the precision determinations of CKM elements using
inclusive semileptonic decays: e.g., for $|V_{cb}|$ a relative uncertainty of
less than 2\% has been achieved based on OPE methods.

To validate and even improve on such accuracy one has to analyze all potential
sources of uncertainties. The OPE is widely viewed as yielding an expansion in
inverse powers of the $b$ quark mass $m_b$ for sufficiently inclusive
quantities. However, in $b \to c$ transitions a second scale is present that
has a reasonable claim to be considered heavy, namely the charm quark mass
with $m_c \gg \Lambda_{\rm QCD}$. It is then natural to integrate out the
charm quark as well. This is usually done for both quarks
at the same scale, which means that the ratio $m_c / m_b$ is
treated as a number of order unity.

On the other hand, numerically we have $m_c^2 \sim m_b \Lambda_{\rm QCD}$ and
hence $m_b \gg m_c \gg \Lambda_{\rm QCD}$. This suggests an alternative where
one integrates out the beauty quark in a first step, while leaving the charm
quark still a dynamical quark; the latter is subsequently removed in a
second step. In this procedure one has an intermediate theory still containing
charm quarks as {\em dynamical} entities. Such a description contains
dimension-six (or higher) operators of the schematic form
\begin{equation}
{\cal O}_{\rm IC} = (\bar{b}_v \Gamma^\dagger c) \, (\bar{c} \Gamma b_v )
\end{equation}  
where $b_v$ is the (now) nonrelativistic $b$ quark field and $\Gamma$ a Dirac
matrix depending on the process under consideration; it may contain
derivatives yielding higher-dimension operators. It has been discussed in some
detail in
\cite{Bigi:2005bh,Breidenbach:2008ua} that these operators 
match onto a contribution proportional to the Darwin term $\rho_D^3$ upon
integrating out the charm quark. Yet new subtleties emerge at this point. This
contribution contains an infrared (IR) sensitive piece proportional to
$\ln{m_b/m_c}$, which diverges for $m_c \to 0$. It is generated in the 
two-step procedure by the RG running between $m_b$ and $m_c$, while it
appears simply in the coefficient function when integrating out bottom and
charm together. In fact, it has been shown in \cite{Bigi:2005bh} that at
higher orders in the OPE even  inverse powers of $m_c$ do appear which
have a stronger infrared sensitivity.

Based on a superficial similarity with various non-valence nonperturbative
charm effects previously discussed in the literature \cite{brodsky}, these
infrared sensitive terms have been dubbed ``Intrinsic Charm'' (IC) in
inclusive $B$ decays. They were largely treated as a factor contributing to
the theoretical uncertainty from higher-order effects \cite{Bigi:2005bh}. In
this paper we show how the seemingly different reasonings given in
\cite{Bigi:2005bh,Breidenbach:2008ua} are equivalent qualitatively as well as
quantitatively and perform a systematic and more complete treatment of these
terms.  It turns out that the conventional OPE yields a combined expansion in
powers of $1/m_b^k \!\times \!1/m_c^l$ with $l>0$ (in fact, starting with
$l\!=\!2$ at tree level, without extra gluon loop) appearing first for
$k\!=\!3$. This gives rise to a quantitative subtlety as well: Since as
mentioned above $m_c^2 \sim m_b \Lambda_{\rm QCD}$, one has to account for the
terms $1/(m_b^3 m_c^2)$ in order to complement the full set of $1/m_b^4$
corrections.

The remainder of the paper is organized as follows. In the next section we
perform the standard procedure and integrate out bottom and charm together in
a similar way at a scale below $m_c$ and trace the origin of the intrinsic 
charm contributions
in this approach. In Sect.~\ref{ALT} we perform the two-step procedure;
i.e., first we integrate out just the bottom quark initially leaving charm as
a dynamical quark in the theory before integrating it out as well at a lower
scale. As expected these two ways yield the same results equivalent to what
has been discussed in \cite{Bigi:2005bh}. We provide a more consistent and
concise derivation of the relation between (generalized) Weak Annihilation and
the four-quark operators than it was done in Ref.~\cite{WA}.
In Sect. \ref{HIGHER} we discuss
explicitly the dimension-eight operators that yield $1/m_b^3 \!\times\! 1/m_c^2$
terms at tree level qualitatively as well as quantitatively, confirming the
results of \cite{Bigi:2005bh}. We summarize the lessons learnt on QCD's
non-perturbative dynamics, the implementation of the OPE and on the
uncertainties in the extraction of $V_{cb}$  in
Sect.~\ref{OUT}. An estimate of the potential scale of Weak Annihilation 
(WA) for $b\tto u$ is
given there as well.

\section{The standard OPE for \boldmath $B \tto X_c \,\ell \bar{\nu}_\ell\,$
revisited } 
\label{FUNDA}

One starts by considering the doubly differential rate for $B \tto X_c \,\ell
\bar{\nu}_\ell$ 
\begin{equation} \label{doubly}
\frac{d^2 \Gamma}{d (v\!\cdot\! p) dp^2} = \frac{G_F^2|V_{cb}|^2}{24\pi^3}
\sqrt{(v\!\cdot\!p)^2 \!-\! p^2} \; 
\theta\!\left((v\!\cdot\! p)^2 \!-\! p^2\right) 
\theta\left(m_b\!-\!v\!\cdot\! p\right) \theta\left(m_b^2\!+\!p^2\!-\!2m_b 
v\!\cdot\! p\right)
\: L _{\mu
	\nu} W^{\mu \nu}(p)
\end{equation}
where $p = m_b v - q$ with $q$ denoting momentum transfer to the lepton pair
and $v$ the $B$ meson velocity. $L_{\mu \nu}$ is the leptonic tensor with
\begin{equation} \label{LepTen}
	L_{\mu \nu}(p) = - m_b^2 [g_{\mu \nu} - v_\mu v_\nu] + m_b [2 g_{\mu
	\nu} (v\cdot p) - v_\mu p_\nu - p_\mu v_\nu ] - [g_{\mu \nu} p^2 -
	p_\mu p_\nu ]
\end{equation}
and $W^{\mu \nu} $ the hadronic counterpart given by
\begin{equation} \label{HadTen}
	2M_B W_{\mu \nu} (p) = \int d^4 x \, \exp(i p\cdot x) \langle B(v) |
	\bar{b}_v (x)  {\Gamma}_\nu c(x) \, \, \bar{c}(0) \Gamma_\mu
	b_v(0) | B(v) \rangle
\end{equation}
with $b_v$ the `rephased' (nonrelativistic) $b$ quark field,  
$b_v (x) = \exp({im_b (v\!\cdot\! x)})\: b(x)$ and 
$\Gamma_\mu = \gamma_\mu (1-\gamma_5)$.

\subsection{General argument}

Here we scrutinize the standard way of setting up the OPE for this inclusive
process treating both the bottom and the charm quarks as heavy 
and neglecting a hierarchy of scales between the bottom and the charm masses. 
We remove both from the dynamical degrees of freedom in one step at a scale 
$\mu$ below the charm mass. No operators with charm fields can then 
emerge in the OPE, and only operators with nonrelativistic $b$ quarks and their
(covariant) derivatives appear.\footnote{We do not discuss here operators
involving light quarks appearing additionally at ${\cal O}(\alpha_s)$ level, their contribution is small.}

The OPE is constructed by contracting the two charm quarks into (the imaginary
part of) a propagator, while the beauty quark field is replaced by its
expansion in inverse powers of $m_b$. Formally expanding the bi-local operator
in (\ref{HadTen}) in local operators yields, upon integration, the well known
Heavy Quark Expansion (HQE). The expansion can be done by external field
methods particularly economical at tree level, see
e.g.~\cite{Dassinger:2006md}. This yields ($b_v \equiv b_v (0)$)
\begin{align}
	2M_B W_{\mu \nu}(p) &= \langle B(v) | \bar{b}_v  {\Gamma}_\nu
	(\fmslash{p}+m_c) \Gamma_\mu b_v | B(v) \rangle \, \delta_+(p^2 - m_c^2)
	\nonumber \\ &+ \langle B(v) | \bar{b}_v  {\Gamma}_\nu
	(\fmslash{p}+m_c) (i \fmslash{D}) (\fmslash{p}+m_c) \Gamma_\mu b_v |
	B(v) \rangle \, \delta^\prime_+ (p^2 - m_c^2) \nonumber \\ &+ \frac12
	\langle B(v) | \bar{b}_v  {\Gamma}_\nu (\fmslash{p}+m_c) (i
	\fmslash{D}) (\fmslash{p}+m_c) (i \fmslash{D}) (\fmslash{p}+m_c)
	\Gamma_\mu b_v | B(v) \rangle \, \delta^{\prime \prime}_+ (p^2 - m_c^2)
	\nonumber \\ &+ \cdots \nonumber \\ &= \sum_{n=0}^\infty
	\frac{1}{n!}\langle B(v) | \bar{b}_v  {\Gamma}_\nu
	(\fmslash{p}+m_c) \left[ \vphantom{\int} (i \fmslash{D})
	(\fmslash{p}+m_c) \right] ^n \Gamma_\mu b_v | B(v) \rangle \,
	\delta^{(n)}_+ (p^2 - m_c^2)\,.  \label{HadTen1}
\end{align}
The main challenge in evaluating higher order contributions lies in
identifying the independent hadronic parameters controlling the `string'
of matrix elements
\begin{equation} \label{nonpert}
 \langle B(v) | \bar{b}_v (iD_{\mu_1}) (iD_{\mu_2}) \cdots (iD_{\mu_n}) b_v |
 B(v) \rangle
\end{equation} 
entering at order $n+1$. It should be noted that these quantities do not
depend on $p$.

Contracting the hadronic and leptonic tensors for the pseudoscalar $B$ mesons
yields a function of $(v\cdot p)$ and $p^2$ of the general form
\begin{equation}  \label{LW}
L _{\mu \nu} W^{\mu \nu} = \sum_{n=0}^\infty \sqrt{(v\cdot p)^2 - p^2} P_n(
(v\cdot p), p^2) \delta_+^{(n)} (p^2 - m_c^2)
\end{equation}
with $P_n( v\cdot p)$ denoting a polynomial in $v\cdot p$. The analysis so far
referred to a fully differential distribution in general kinematics. To
proceed to the inverse mass expansions we need to consider partially integrated
probabilities. 
 
The integration over the variable $v\cdot p$ has the limits $\sqrt{p^2} \le
v\cdot p \le (m_b^2 + p^2) / (2 m_b)$, which yields terms logarithmic in $p^2$
from the lower end of integration.  Focusing on these logarithms we get for
$l\!=\!0, 1, ...$
\begin{align}  \label{master}
	\int_{\sqrt{p^2}} d(v\cdot p) \, \sqrt{(v\cdot p)^2 - p^2} \, (v\cdot
	p)^{2l} &= C_l \, \, (p^2)^{l+1} \ln\left(\frac{p^2}{m_b^2} \right) +
	\cdots \,, \quad 
C_l=\frac{\Gamma(l\!+\!\mbox{$\frac{1}{2}$})}{4\sqrt{\pi}\,\Gamma(l\!+\!2)}\;\;
\\ 
\int_{\sqrt{p^2}} \!d(v\cdot p) \, \sqrt{(v\cdot p)^2 - p^2}
	\, (v\cdot p)^{2l\!+\!1} \!&= 0 + \cdots 
\nonumber
\end{align}  
where the ellipses denote polynomial terms in $p^2$, and the 
coefficients $C_l$ are simple fractions: 
$C_0 \!=\! 1/4$, $\,C_1 \!=\! 1/16$, $\,C_2 \!=\! 1/32$, $\,C_3 \!=\! 5/256$ ...

These logarithms are the source of the IR sensitivity of the coefficient
functions to the charm mass in this approach. This becomes manifest when they
are combined with the derivatives of the $\delta$-function in Eq.(\ref{LW}): 
\begin{align} \label{DerivDelta}
(p^2)^k \ln{p^2}\; \delta^{(k)} (p^2 \!-\! m_c^2) &=
(-1)^k \,k!\: \ln{m_c^2}\; \delta (p^2 - m_c^2) + \cdots
\\ (p^2)^k \ln{p^2}\; \delta^{(n)} (p^2 - m_c^2) &=
(-1)^{n\!-\!k\!-\!1} \,k!(n\!-\!k\!-\!1)!\: \left(\frac{1}{m_c^2}
\right)^{n-k} \delta (p^2\!-\! m_c^2) \quad \mbox{for } n>k
\end{align}
where $k$ is some integer power and the ellipses 
point to less singular terms as $m_c \to 0$. Note that at
the lower limit of integration in Eq.~(\ref{master}) we have $p_0^2\!=\!p^2$,
i.e.\ $\vec{p}\tto 0$. This is the infrared regime for the charm quark in the
final state.

The leading IR sensitive terms in the integrated rates arise from only the
leading term in the leptonic tensor
\begin{equation} \label{LepTenLead}
	L_{\mu \nu}^\text{lead}  = - m_b^2 [g_{\mu \nu} - v_\mu v_\nu] 
\end{equation}
already in the expression for the differential rate. For the terms containing
the vector $p$ would lead to powers of $v\cdot p$ and $p^2$. To obtain 
non-leading in $1/m_b$ terms, we consider the subleading terms in the leptonic tensor
\begin{align}
	L_{\mu \nu}^{\rm \,sub} &= m_b [2 g_{\mu \nu} (v\cdot p) - v_\mu
	p_\nu - p_\mu v_\nu ] \label{LepTenSubLead} \\ L_{\mu
	\nu}^\text{\rm \,subsub} &= - [g_{\mu \nu} p^2 - p_\mu p_\nu
	]\,.\label{LepTenSubSubLead}
\end{align}
On the other hand, the $n^{th}$ term in the sum (\ref{HadTen1}) for the
hadronic tensor contains
\begin{equation}
	P_n \propto  {\Gamma}_\nu (\fmslash{p}+m_c) \gamma_{\mu_1}
	(\fmslash{p}+m_c) \gamma_{\mu_2} \cdots (\fmslash{p}+m_c)
	\gamma_{\mu_n} (\fmslash{p}+m_c) \Gamma_\mu
\end{equation}
which yields upon contraction with the leptonic tensor (\ref{LepTenLead}) and
with the nonperturbative matrix elements (\ref{nonpert}) a contribution of the
form (see (\ref{LW}))
\begin{equation}
	P_n = \sum_{ijl} a_{ijl} \, \, (v\cdot p)^i (p^2)^j (m_c^2)^l \quad
	\mbox{with} \quad i+2j+2l = n+1 \,.
\end{equation}
Note that due to the purely left-handed structure of the current $\Gamma$, we
can have only even powers of $m_c$ here. Furthermore, if $n$ is even, $i$
necessarily is odd (and vice versa). Hence from (\ref{master}) we conclude
that IR sensitive terms can appear only if $n$ is odd, which means that the number of
covariant derivatives in the hadronic matrix element has to be odd as
well. Therefore all operators which contribute to intrinsic charm have to
match onto partonic matrix elements with at least one gluon. In turn, it
implies that there is no intrinsic charm contribution to operators of the form
$\langle (\vec{k}^2)^n \rangle$ where $\vec{k}$ are the spatial components of
the residual $b$ quark momentum.

Now we can trace how singular terms actually emerge. 
Putting everything together, we get for $i = 2m$ 
\begin{align}
	& \int d(v\cdot p) \sqrt{(v\cdot p)^2 - p^2} P_n( (v\cdot p), p^2)
	\delta^{(n)} (p^2 - m_c^2) \\ & = \sum_{ijl} a_{ijl} \, \, \int
	d(v\cdot p) \sqrt{(v\cdot p)^2 - p^2} \left[ (v\cdot p)^i (p^2)^j
	(m_c^2)^l \right] \delta^{(n)} (p^2 - m_c^2) \\ & \Longrightarrow
\sum_{ijl} C_l\, a_{ijl}
	\, \, (p^2)^{m+j+1} (m_c^2)^l \ln\left(\frac{p^2}{m_b^2} \right)
	\delta^{(n)}(p^2 - m_c^2) \; .
\end{align}
Terms with odd $i$, on the other hand, do not contain a logarithm. Note that
we have $2 m + 2 j + 2 l = n+1$, which can be satisfied at any odd $n$. Thus
we arrive at 
\begin{align} &\int d(v\cdot p) \sqrt{(v\cdot p)^2 - p^2} P_n(
(v\cdot p), p^2) \delta^{(n)} (p^2 - m_c^2) \\ &
\stackrel{\scalebox{.5}{sing}}{=} \sum_{ijl} 
C_l\, a_{ijl} \, \,
(p^2)^{(n+3-2l)/2} (m_c^2)^l \ln\left(\frac{p^2}{m_b^2} \right) \delta^{(n)}
(p^2 - m_c^2) \,.
\end{align}
Hence IR sensitive terms can appear starting at $n=3$ with the logarithmic
dependence on $m_c$ of the Darwin term. For $n > 3$ we get from these terms a
tree level contribution to the total rate of the form
\begin{equation}
	\Gamma_n \propto \frac{1}{m_b^3} \left(\frac{1}{m_c^2}
	\right)^{(n-3)/2} \quad \mbox{with } n = 5,7,9 ... \; ,
\end{equation} 
i.e., even a powerlike singularity for $m_c \tto 0$. 

\subsection{Explicit expressions for \boldmath $n\!=\!0\,$ through $\,4\,$ in the
total width}

For $n \!=\! 0$ the leading term reads 
\begin{equation}
	P_0^\text{lead} = \frac32 m_b^2 (v\cdot p)
\end{equation} 
which is an odd power of $v\cdot p$ and hence does not lead to IR sensitive
terms as just explained. One should note that the subleading contribution to
the leptonic tensor $L_{\mu \nu}$ yields -- upon contraction with
the partonic hadronic tensor -- a (subleading) term of the form
\begin{equation}
	P_0^{\rm \,sub} = -2 m_b (v\cdot p)^2 - p^2 m_b \; ; 
\end{equation} 
it leads to a contribution of the form $m_c^4 \ln (m_c^2)$ in the phase space
factor of the partonic total rate. We will focus eventually
on those novel terms $\propto \!1/m_c^k$ that arise in leading order in $1/m_b$;  
to low orders in 
$\Lam$ it is easy to keep as well the terms stemming from the subleading 
pieces in $L_{\mu \nu}$.

The next term with $n=1$ is given by 
\begin{equation}
	P_1= \frac{\mu_G^2-\mu_\pi^2}{12 m_b} \left(5 p^4+7 m_c^2 p^2-20
	(v\cdot p) ^2 p^2 - 10 m_c^2 (v\cdot p)^2\right)\,, 
\end{equation} 
where the hadronic tensor is contracted with the subsubleading part of the
leptonic tensor $L_{\mu \nu}^{\rm \,subsub}$.  This yields again an $m_c^4
\ln (m_c^2)$ terms upon integration over the phase space. 

For $n\!=\!2$ the leading term of the leptonic tensor again contains only odd
powers of $(v \cdot p)$ which do not generate any logarithms.

At $n=3$ the IR sensitive contribution is the Darwin term. Explicitly, we have
\begin{equation}
 	\langle B(v) | \bar{b}_v (iD_\alpha) (iD_\gamma) (iD_\beta) b_v | B(v)
 	\rangle = \frac16 M_B \rho_D^3 (g_{\alpha \beta} - v_\alpha v_\beta)
 	v_\gamma (\fmslash{v} + 1) + \cdots
\end{equation}
from which we obtain  
\begin{equation}
	P_3^{\rm \,Dar} = -\frac{\rho_D^3}{12} m_b^2 \left(3(p^2 \!-\! m_c^2)^2
+8( v\cdot p)^4-8 p^2 (v\cdot p)^2\right) .
\end{equation} 
Upon integration over $(v\cdot p)$ we arrive at terms with three types of prefactors:
\begin{equation}
(p^2)^3 \ln{\left(\frac{p^2}{m_b^2} \right)} \; , \quad m_c^2 (p^2)^2
\ln{\left(\frac{p^2}{m_b^2} \right)}\; , \quad m_c^4 p^2
\ln{\left(\frac{p^2}{m_b^2} \right)} \; .
\end{equation}
They have three derivatives with respect to $p^2$ from the 
$\delta ^{(3)}$-function,  and hence the first term 
yields a $\ln (m_c^2)$ factor which is the
first infrared sensitive contribution. In the other two terms explicit factors
of $m_c^2$ kill the infrared singularity and they thus remain finite for
$m_c \tto 0$.  It is straightforward to check that in this way we end up with 
the correct prefactor for the infrared log in the Darwin contribution.

The terms with $n\!=\!4$ create again only odd powers of $(v \cdot p)$ and would
yield infrared singularity for the $1/m_b$-subleading piece. Finally at $n\!=\! 5$ 
the following nine structures arise 
\begin{eqnarray}
P_5 \propto && (v\cdot p)^6\; , \quad (v\cdot p)^4 p^2 \; ,
\quad (v\cdot p)^4 m_c^2\; ,  \quad  (v\cdot p)^2 (p^2)^2\; ,  
\quad (v\cdot p)^4 m_c^4 \; ,  \\  
 && (v\cdot p)^2 p^2 m_c^2 \; , \quad (p^2)^3\; , \quad (p^2)^2 m_c^2\; ,
 \quad p^2 m_c^4 \nonumber
\end{eqnarray}
Upon integration over $(v\cdot p)$ we obtain terms of the form
\begin{eqnarray}
(p^2)^4 \ln\left(\frac{p^2}{m_b^2} \right) \; , \quad (p^2)^3 m_c^2
\ln\left(\frac{p^2}{m_b^2} \right) \; , \quad (p^2)^2 m_c^4
\ln\left(\frac{p^2}{m_b^2} \right) \; , \quad p^2 m_c^6
\ln\left(\frac{p^2}{m_b^2} \right)
\end{eqnarray}
coming with five derivatives of the $\delta$-function.  All thus yield
contributions of order $1/m_c^2$ in the total rate. They will be addressed in
detail in the next section. A similar consideration extends in a
straightforward way to higher orders where $n\!>\!5$ emerge. These would
generate terms inversely proportional to even larger powers of inverse charm
mass. \vspace*{2pt}

As a first resum\'e we state that IR sensitive contributions -- i.e., those
singular for $m_c \to 0$ -- unequivocally arise from the lower end of the
integration over $v\cdot p$ (i.e.\ $\vec{p}\tto 0$) due to the presence 
of the non-analytic factor
$\sqrt{(v \cdot p)^2 \!-\! p^2}=|\vec{p}\,|$ in the integrand. For the dimension-six Darwin
term they are of the form $\ln{(m_b^2/m_c^2)}$. Higher-dimension
contributions exhibit an even stronger singularity, viz.\ powers of $1/m_c$. Without
extra gluon loops this happens first for dimension eight.

The discussion above makes it clear that such IC effects emerge for the fully
integrated width as well as for higher moments of the distributions. However the
strength of the infrared singularity and the order in the heavy quark
expansion it first emerges generally depends on the kinematic observable.

\section{ \boldmath $m_b \gg m_c \gg \Lambda_{\rm QCD}$:   
charm as a dynamical quark} 
\label{ALT}

In this section we shall consider an alternative way to describe IC
effects. We now choose to integrate out the `heavy' degrees of freedom only
above the scale $m_c$. This leaves the charm quark as a dynamical entity, much
in the same way as would be required for light quarks in QCD, e.g.\ in $b\tto
u\,\ell\nu$. The main difference that emerges is that, for inclusive decays,
we have now to include four-quark operators explicitly containing charm quark
fields.

The charm quarks in $b$ decay can act both as a hard and a soft degree of freedom. 
The hard component is
treated the same way as described in the previous section. The matrix elements
of the four-quark operators contain the ``soft'' part of the still dynamical
quarks, which for now we treat nonperturbatively. Accordingly, we write the
original QCD product of currents as 
\begin{eqnarray} \label{3.1} 
 &&\matel{B}{\bar{b} (x)  {\Gamma}_\nu c(x) \, \bar{c}(0) \Gamma_\rho
 b(0)}{B} \\ && = \nonumber \matel{B}{\bar{b} (x)  {\Gamma}_\nu \,
 \langle c(x) \bar{c}(0) \rangle \, \Gamma_\rho b(0)} {B}_{>\mu} +
 \matel{B}{\bar{b} (x)  {\Gamma}_\nu c(x) \, \bar{c}(0) \Gamma_\rho
 b(0)}{ B }_{<\mu} \; .
\end{eqnarray} 
The product $c(x) \bar{c}(0)$ can be viewed as the Green function of the charm 
quark inside the
external gluon field in the $B$ meson, averaged over the field configurations
present in the meson. This is an exact relation as long as the beauty meson 
has no charm flavor: it is a consequence of the Gaussian
integration over the quark fields. This applies to the l.h.s. as well as to each of
the two terms on the  r.h.s. The decomposition on the r.h.s. merely reflects 
the different treatment used to describe these terms.

The first term corresponds to the `perturbative' (in $1/m_c$) calculation of
the previous section. The second term has to be added now, since the charm
quark is still a dynamic quark controlled by nonperturbative dynamics. The
role of the intermediate scale $\mu$ is to draw the demarcation between 
the two dynamical regimes. 

Even though the first term in Eq.~(\ref{3.1}) is evaluated in the `direct' way
detailed in the previous section, the result differs due to the introduction of the  cutoff
$\mu$. The precise form of how $\mu$ enters depends on the
concrete way the (hard) separation is implemented. Let us mention that for
tree-level calculations without extra perturbative loops it is sufficient and
convenient to simply integrate the distribution with the constraint 
\beq
0 \le q^2 < (m_b\!-\!\mu)^2.
\label{3.20}
\eeq
The upper bound above effectively
introduces the separation scale in Eq.~(\ref{3.1}) when the correlator is
integrated over the phase space to obtain the inclusive width or its 
moments.

Evaluating the first term in Eq.~(\ref{3.1}) in the previous section gave
contributions IR sensitive to the charm mass. Taking the formal limit $m_c\tto
0$ separately term by term in the expansion
would have yielded divergent
expressions. With a nonzero $\mu \!\gg\! \Lam$ this changes: all individual
terms remain regular by themselves at $m_c\tto 0$. For the role of the
infrared regulator is now taken over by the Wilsonian cutoff.

This is evident on general grounds: the terms IR-singular for $m_c\tto 0$ came
only from the soft charm configuration with momentum $p \lsim m_c$ -- the
domain now excluded. Alternatively, this can be traced explicitly in the
formalism of Sect.~2. For instance, with the constraint of Eq.~(\ref{3.20})
the lower limit of integration in $(v\cdot p)$ rises from $\sqrt{p^2}$ to
$\frac{p^2+2m_b \mu -\mu^2}{2m_b}\simeq \mu$; therefore
$\log{\frac{m_b^2}{p^2}}$ in Eq.~(\ref{DerivDelta}) turns into
$\log{\frac{m_b^2}{\mu^2}}$. All the integrals become analytic functions of
$m_c$ at $m_c \!\ll\!\mu$; the logs and inverse powers of $m_c$ are replaced
by those of the cutoff mass $\mu$.

The second term in the r.h.s.\ of Eq.~(\ref{3.1}) has a smooth $m_c\tto 0$
limit. Although it may have soft ``chiral'' singularities when both $m_c$ and
one of the light quarks become massless, the expectation values should remain
finite; only higher derivatives with respect to the charm mass may have
singularities if $m_u$ or $m_d$ vanish. We note that this expectation value,
being regularized in the ultraviolet, is well-defined and these conclusions
hold with no reservations.

We may briefly discuss at this point the parametric dependence of the total
semileptonic width $\Gamma_{\rm sl}(b\tto c)$ on $m_c$ when the latter is
small. As long as $m_c \!\gg\! \Lam$ holds, we have non-analytic
nonperturbative terms at order $1/m_b^3$ scaling like 
$\frac{\Lam^3}{m_b^3} \ln{\frac{m_b^2}{m_c^2}}$, $\frac{\Lam^3}{m_b^3}
\left(\frac{\Lam^2}{m_c^2}\right)^k$ with $k\!>\!0$ 
(in general, odd powers of $m_c$ also emerge). Each of these terms separately
are singular at $m_c \tto 0$. The leading term is driven by the Darwin
expectation value; it represents an IR singularity in $m_b/m_c$ which, in
principle, is observable, at least at large $m_b$ and sufficient accuracy.

The expansion in $\Lam/m_c$ makes, however, sense only as long as charm
remains heavy on the scale of QCD dynamics. At lower $m_c$ the successive
terms with higher $k$ would formally dominate. The whole function of $m_c$
stabilizes at $m_c \lsim \Lam$ and approaches a finite value at 
$m_c\tto 0$. A model for such a behavior can be given, for instance, by
\beq
\frac{\rho_D^3}{m_b^3} \,\ln{\frac{m_b^2}{m_c^2\!+\!\Lambda^2\!\!}}
\label{3.26}
\eeq
(here $\Lambda$ is a strong interaction mass scale parameter of the order of
$\Lam$) which, expanded in $1/m_c^2$, would yield the whole series in
$1/m_c^2$. The actual coefficients for the $1/m_c^{2k}$ terms may, of course,
be different, and they can be calculated in the OPE along either road.

Returning to the OPE analysis proper, 
we can bridge the two ways of accounting for the nonperturbative charm effects
by looking at the $\mu$-dependence of both terms in Eq.~(\ref{3.1}) (more
accurately, when it is integrated to obtain the inclusive probability). Their
sum must be $\mu$ independent which provides useful relations. The following
note should be kept in mind. 

The form of the $\mu$-dependence is determined by the regularization
scheme. In the Wilsonian procedure with a hard cutoff the leading 
log dependence is accompanied by power terms. The resulting dependence is
qualitatively different for $\mu \!\ll\! m_c$ and for $\mu \!\gg\! m_c$. \\
(i) When
$\mu$ is taken small compared to $m_c$, it enters only as a small power
correction, $\sim \!(\mu/m_c)^l$; in the formal limit $\mu\tto 0$ ($m_c
\!\gg\!\Lam$ fixed) the last term would vanish, the whole correction to the
width is then given by the first term. Raising $\mu$ up to $m_c$ and above 
moves the correction to
the width from the first term to the second.  The calculation of
Sect.\,2 represents the evaluation of the last term contribution in the 
$\Lam/m_c$-expansion for $\mu \!\gg \!m_c$. Such a calculation makes 
sense only as long as charm is sufficiently heavy.\\
(ii) At $\mu \! \gg \! m_c$ the situation is different. Beyond the leading
term $O_D \ln{\frac{\mu^2}{m_c^2}}$ the $\mu$ dependence is
suppressed by powers of $m_c^2/\mu^2$. In the second term in Eq.~(\ref{3.1})
these appear as a residual dependence of the high-dimension terms on the
ultraviolet cutoff for integrals  intrinsically convergent at
the scale $m_c$. In the first term it shows up as now small coefficient
functions of higher-dimension operators -- which would normally be saturated
at soft charm configurations $p\sim m_c$ -- proportional to $1/m_b^3\,
1/\mu^{2k}$, instead of $1/m_b^3 \,1/m_c^{2k}$ without a cutoff. It is then
convenient to assume $\mu \!\gg \!m_c$ and neglect these terms altogether.

A short comment is in order on how the separation would look like in the 
often adopted dimensional regularization scheme (DR)
where no powerlike dependence on renormalization scale ever arises. 
For small $m_c$  the only $\mu$-dependence within DR enters through 
$\ln{\mu}$ in the 
Darwin operator. In such a scheme, however, charm quarks must be treated as
massless;
the dynamic four-quark operators have to be renormalized in the 
UV likewise in the way of DR. The requirement to set $m_c\!=\!0$ then is
rather clear. For keeping $m_c$ finite would
destroy naive DR in a direct calculation: it yields a 
finite result for the first term around $D\!=\!4$ because the potential 
IR singularity at
$D\!=\!4$ is regularized by a non-zero charm mass. As a result, with
$m_c\!\ne\!0$ calculating the first term  in Eq.~(\ref{3.1}) with DR  
precisely reproduces the 
total contribution of Sect.~2. Consequently, in DR at $m_c\!\ne\!0$ the last
term in Eq.~(\ref{3.1}) should be considered as vanishing regardless of
the concrete value of the charm mass; this is not a very physical 
result, at least for a relatively light quark. 

This is a rather typical feature of dimensional regularization. DR may provide
a convenient highly efficient technical tool to analyze the case of massless
final state quark or when a non-zero charm mass can be neglected. However, it would
require additional matching procedure, usually order by order in $m_c$ if the
charm mass dependence is important.
\vspace*{4pt}

In logarithmic terms -- as for the Darwin operator in the integrated width --
the charm mass plays the role of a renormalization point. Therefore, as
discussed in Ref.~\cite{Breidenbach:2008ua}, the infrared dependence on $m_c$
in the `conventional' calculation along the `first' road must match the UV
dependence of the corresponding four-quark expectation value given by
$\rho_D^3 \ln{\frac{\Lambda_{\rm UV}^2}{m_c^2}}$. A similarly dual description
applies also for the terms scaling like inverse powers of $m_c$. They can be
calculated conventionally assuming charm to be heavy following the route of
Sect.~2. Alternatively, they can be obtained as the corresponding pieces of
the four-quark expectation value (normalized at $\mu \!\gg \!m_c,
\Lam$). The results over either road must be identical whenever one is in
the domain where the expansion can be applied. The first road (without
implementing a cutoff) is, of course, justified only for $m_c\!\gg
\!\Lam$. The second route is formally valid for an arbitrary hierarchy between
$m_c$ and $\Lam$ -- however, we do not have the means to calculate the
expectation value through gluon operators without charm fields when charm
becomes light.

In order to make the above mentioned correspondence explicit, we will address
the $1/m_b^3 \, 1/m_c^n$ terms where at tree level only even $n$ emerge. To
this end, we consider the contribution of the second term in Eq.~(\ref{3.1})
involving the explicit charm quark operators. Inserting it into the hadronic
tensor we get
\begin{align}
	2M_B W_{\mu \nu}^{\rm (IC)} &= \int d^4 x \, \exp(i p\cdot x) \langle B(v)
	| \bar{b}_v (x)  {\Gamma}_\nu c(x) \, \, \bar{c}(0) \Gamma_\mu
	b_v(0) | B(v) \rangle |_\mu \\ \nonumber &= \int d^4 x \, \exp(i
	p\cdot x) \langle B(v) | \bar{b}_v (0)  {\Gamma}_\nu c(0) \,
	\, \bar{c}(0) \Gamma_\mu b_v(0) | B(v) \rangle |_\mu \\ \nonumber
	&\quad + \int d^4 x \, \exp(i p\cdot x) x_\alpha \langle B(v) | [
	\partial^\alpha ( \bar{b}_v (0)  {\Gamma}_\nu c(0))] \, \,
	\bar{c}(0) \Gamma_\mu b_v(0) | B(v) \rangle |_\mu + \cdots
\end{align}  
where the ellipses denote operators of dimension eight and higher. 
Extra derivative for (renormalized) expectation values can bring in a factor
of $m_c$ or $\mu$ at most; it can be traced that powers of $x_\alpha$
translate into powers of $1/m_b$. Therefore, the higher terms in this
expansion yield higher powers in the $1/m_b$ expansion. We then can
focus on the first term which is the conventional $D\!=\!6\,$ four-quark
operator. The corresponding hadronic tensor reads 
\bea
\nonumber
2M_B W_{\mu \nu}^{\rm (IC)} &\msp{-5}=\msp{-5}& (2 \pi)^4 \delta^4 (p) \, \matel{B}{\bar{b} (0)
 {\Gamma}_\nu c(0) \,  \bar{c}(0) \Gamma_\mu b(0)}{B}_\mu \\ 
& + & (2
\pi)^4 \left(-i\frac{\partial}{\partial p_\alpha}\right)\delta^4 (p) \;
\matel{B}{\partial_\alpha \bar{b} 
 {\Gamma}_\nu c(0) \,  \bar{c}(0) \Gamma_\mu b(0)}{B}_\mu \, + ...
\label{ICTen} 
\eea 
and we retain only the first term. Note that the $\delta$-function projects
out the leading term of the leptonic tensor (\ref{LepTenLead}). Furthermore,
this contribution is localized at $p = 0$ -- hence at $p^2 \!=\!  0$ and $v
\!\cdot \! p \!=\! 0$, in agreement with the findings of the last
section. 
As a consequence, step-functions in Eq.~(\ref{doubly}) become 
superfluous. The last relation
completing the arithmetic part is 
\beq
\int \!{\rm d}(v\!\cdot\! p)\,{\rm d}p^2\,\sqrt{(v\!\cdot\! p)^2\!-\!p^2}
\;\;\delta^4 (p) =\frac{1}{2\pi}\;.
\label{3.52}
\eeq

Omitting QCD corrections, the relevant diagrams for calculating 
$W_{\mu \nu}^{\rm (IC)}$ in Eq.~(\ref{ICTen}) are the one loop diagrams
involving the charm-quark loop with an arbitrary number of external
gluons. These diagrams have been considered already in \cite{Bigi:2005bh}. It
is advantageous to first perform a Fierz rearrangement of the four quark
operator according to 
\begin{equation}  \label{fourquark} 
	\bar{b}_{\,}  {\Gamma}^{\nu\!} c \, \, \bar{c}_{\,} \Gamma^\mu b =
	-\frac12 \bar{b}_{\alpha}  {\Gamma}_\rho b_{\beta} \, \,
	\bar{c}_\beta \Gamma_\sigma c_\alpha \left[-i \epsilon^{\mu \nu \rho
	\sigma} + g^{\sigma \mu} g^{\rho \nu} + g^{\sigma \nu} g^{\rho \mu} -
	g^{\rho \sigma} g^{\mu \nu} \right], 
\end{equation}
where $\alpha$, $\beta$ are color indices, and the minus sign comes from
anticommutativity of the quark fermion fields.

We may construct 
the charm mass expansion of this charm loop in the external gauge field
and take the average over the $B$-meson state. There are a few
subtleties related to this procedure, since the two charm quark operators
are taken at coinciding space-time points.  As in \cite{Bigi:2005bh} we may
start from the time-ordered product of two charm quark operators at displaced
points, which amounts to consider the conventional charm propagator in an 
external field. This propagator is generally gauge dependent, yet in the end this is
compensated by the same displacement in the $b$-quark fields.  For
constructing the short-distance expansion of the Green function the fixed-point
gauge is convenient.

\begin{figure}[!h]
\centerline{
\includegraphics[width=150pt]{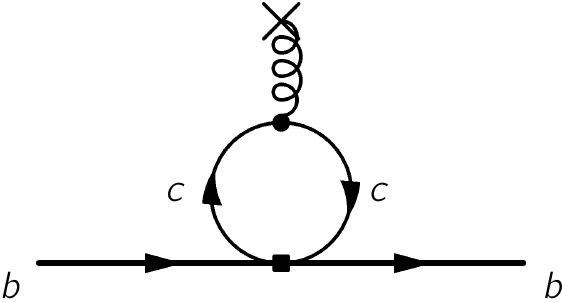} \hspace*{30pt}
\includegraphics[width=150pt]{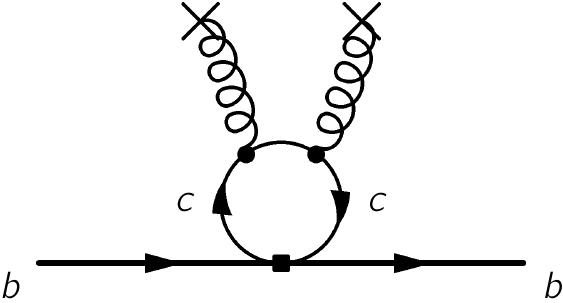}}
\caption{Diagrams illustrating calculation of the charm loop in the
external field. The wavy lines generically reflect insertions of the external gluon
field.}
\label{diagrams}
\end{figure}

The limit of coinciding points in the charm Green function is formally
divergent, yet gauge independent.  Subtracting the free Green function (this
piece is accounted for in the purely partonic width) we end up with a mild log
divergence present in the vector current, proportional to $[D_\mu, G_{\mu\nu}]$:
\begin{equation}  
	\aver{\bar{c}_\alpha \gamma^\nu c_\beta}_A = \frac23
	\frac{1}{(4\pi)^2} \ln \left( \frac{\Lambda_{\rm UV}^2}{m_c^2} \right)
	\left[ D_\kappa \, , \, G^{\kappa \nu} \right]_{\beta\alpha} + \cdots
\label{3.5}
\end{equation}
where $G_{\mu \nu}$ is the gauge field strength tensor; $\Lambda_{\rm UV}$
should be identified with $\mu$ in this context, and the ellipses denote
finite terms to be considered below. As discussed in
\cite{Breidenbach:2008ua}, this ultraviolet-singular log matches onto the 
infrared piece of the conventionally calculated Darwin coefficient
function. The contributions from the axial-vector $\bar{c} c$ current are 
convergent ab initio. Examples of the Feynman diagrams are shown in 
Fig.~\ref{diagrams}.

The lowest finite terms yield $1/m_b^3 \,1/m_c^2$ contributions given by    
\begin{align}
	\langle \bar{c}_\alpha \gamma_\nu \gamma_5 c_\beta \rangle_A &=
	\frac{1}{48 \pi^2 m_c^2} \left(2 \left\{ \left[ D_\kappa , 
	G^{\kappa \lambda} \right], \tilde{G}_{\nu \lambda} \right\} +
	\left\{ \left[ D_\kappa , \tilde{G}_{\nu \lambda} \right] ,
	G^{\kappa \lambda} \right\} \right)_{ \beta \alpha} + \cdots \\
\langle \bar{c}_\alpha \gamma_\nu c_\beta \rangle_A &= \frac{i}{240
	\pi^2 m_c^2} \left( 13 \left[ D_\kappa , \left[ G_{\lambda \nu }
	,  G^{\lambda \kappa} \right] \right] + 8 i \left[ D^\kappa 
	,  \left[ D^\lambda  ,  \left[ D_\lambda  ,  G_{\kappa \nu}
	\right] \right] \right] \right. \\ 
& \qquad \qquad \quad \left. - 4 i
	\left[ D^\lambda  , \left[ D^\kappa ,  \left[ D_\lambda  ,
	 G_{\kappa \nu} \right] \right] \right] \right)_{ \beta \alpha} +
	\cdots \nonumber
\end{align} 
This assumes $\mu \! \gg \!m_c$ and neglecting power terms $\sim
(m_c/\mu)^k$. 
Inserting this into (\ref{fourquark}) we end up with dimension-eight
$\bar{b}...b$ operators with gluon fields; their coefficient functions
compared to the partonic $D\!=\!3$ operator $\bar{b}b$ are proportional to 
$1/m_b^3 \,1/ m_c^2$.

A closer look into the calculations of the charm loop in the external field
reveals that the resulting expressions exactly parallel those in Sect.~2 once
the leading-order approximation Eq.~(\ref{LepTenLead}) is adopted and only
non-analytic terms according to Eq.~(\ref{master}) are retained. This occurs 
before the full integration is performed, when one takes the integral over the
timelike component of the loop momentum by the residues at $p^2\!=\!m_c^2$, for
each power term in the expanded propagator.

In fact, the technique of the loop calculation in the external field itself
allows to derive a number of relations which strongly constrain the form of
the operators which can appear in such an expansion. It has been presented in
detail in the first part of Ref.~\cite{SVZFest}. 
These relations ensure that the result has always the form of multiple
commutators.
This sharpens one observation made already in Sec.~\ref{FUNDA}. There
we noted that the partonic matrix elements of IC contributions necessarily
have to be at least one-gluon matrix elements. From the arguments given in this
section we conclude that intrinsic charm contributions involve only gluon
fields and their derivatives; there will be no derivatives acting on the
beauty quark fields that would generate a dependence on the 
`residual momentum' of the decaying $b$-quark.

We have verified though explicit calculations of the $1/m_b^3\, 1/m_c^2$ terms
that these two ways to calculate $1/m_c$-singular terms yield the same
result for the operator expansion.

At order $1/m_b^3 \, 1/m_c^2$ the result can thus be expressed through
five operators, which are determined by two contributions to the axial vector
current and three contributions to the vector current:
\begin{align} \label{set-of-five} 
	2M_B \tilde f_1 &= \langle B | \bar b_v \Big[i D _\kappa, \big[ i
	D_\lambda, [i D^\lambda ,i G^{\kappa \alpha}]\big]\Big] b_v | B
	\rangle \,\, v_\alpha \\
	2M_B \tilde f_2 &= \langle B | \bar b_v \Big[i D _\lambda, \big[ i
	D_\kappa, [i D^\lambda ,i G^{\kappa \alpha}]\big]\Big] b_v | B \rangle
	\,\, v_\alpha \\
	2 M_B \tilde f_3 &= \langle B | \bar b_v \Big[i D _\kappa, \big [ i
	G_{\lambda \alpha}, i G^{\lambda \kappa} \big] \Big] b_v | B \rangle
	\,\, v^\alpha \\
	2 M_B \tilde f_4 &= \langle B | \bar b_v \big\lbrace \big[ i D ^\rho,
	i G_{\rho \lambda} \big], i G_{\delta \gamma} \big \rbrace (-i
	\sigma_{\alpha \beta}) b_v | B \rangle \nonumber \\ & \qquad \times
	\frac12 \left(g^{\lambda \alpha} g^{\delta \beta} v^\gamma-g^{\lambda
	\alpha} g^{\gamma \beta} v^\delta+g^{\delta \alpha} g^{\gamma \beta}
	v^\lambda \right) \\
	2 M_B \tilde f_5 &= \langle B | \bar b_v \big\lbrace \big[ i D ^\rho,
	i G_{\sigma \lambda} \big], i G_{\rho \gamma}\big \rbrace (-i
	\sigma_{\alpha \beta}) b_v | B \rangle \nonumber \\ & \qquad \times
	\frac12 \left(g^{\sigma \alpha} g^{\lambda \beta} v^\gamma-g^{\sigma
	\alpha} g^{\gamma \beta} v^\lambda+g^{\lambda \alpha} g^{\gamma \beta}
	v^\sigma \right) \,.
\end{align} 
The contributions originating from the axial current yield
spin-triplet operators, while those of the vector current yield spin singlet
operators.

\section{Quantitative estimates of Intrinsic Charm}
\label{HIGHER}

To estimate the IC contributions we follow the lines of \cite{Bigi:2005bh} and
apply the ``ground-state factorization'' approximation to the matrix elements.

The effect on the total rate has been considered already in
\cite{Bigi:2005bh}, yielding a reasonably small contribution. In fact, the
total rate, expressed in terms of the operators given in (\ref{set-of-five})
reads as 
\begin{equation}
	\frac{m_b^3 m_c^2}{\Gamma_0} \;\Gamma\, 
\rule[-15pt]{.3pt}{22pt}_{\raisebox{5pt}{\,$\frac{1}{m_c^2}$}} =
	-\frac32\, \frac{2}{15} \left(-8 \tilde f_1 + 4 \tilde f_2 - 13 \tilde
	f_3 \right) + \frac12\, \frac{2}{3} (-2 \tilde f_4 - \tilde f_5) \,.
\end{equation}
Following  the way to evaluate the expectation values
suggested in Ref.~\cite{Bigi:2005bh} we obtain numerically\footnote{We have found 
a discrepancy with Ref.~\cite{Bigi:2005bh}
in the overall factor for one of the expectation values. It does not produce a
noticeable numerical change for the correction to the width, however.}
\beq
\tilde f_1 \!\approx\! 0.31\,\text{GeV}^5, \quad 
\tilde f_2  \!\approx\!0.25\,\text{GeV}^5, \quad  
\tilde f_3  \!\approx\!0.14\,\text{GeV}^5, \quad 
\tilde f_4  \!\approx\!0.34\,\text{GeV}^5, \quad
\tilde f_5  \!\approx\! -0.40\,\text{GeV}^5 
\label{NumEst}
\eeq
which lead to\,\footnote{Ref.~\cite{Bigi:2005bh} included an additional
phase-space suppression factor for the IC kinematic of
$(1\!-\!m_c/m_b)^2$. Based on the operator analysis we can show that actually
it is absent in the case at hand.}
\begin{equation}
	\delta \Gamma \, 
\rule[-15pt]{.3pt}{22pt}_{\raisebox{5pt}{\,$\frac{1}{m_c^2}$}}
 \approx (0.7\%) \times
	\Gamma_{\rm Parton} \, .
\end{equation}

This numerical estimate should not be considered
bullet-proof. While the individual matrix elements are predicted with
reasonable confidence in their signs and magnitudes, we are faced with a set
of terms with different signs. Thus cancellations will in general occur among
them. Their degree may depend on the numerical accuracy of the applied
ground-state factorization, as well as on the precise values of the
lower-dimension expectation values $\mu_\pi^2$, $\rho_D^3$ and
$\rho_{LS}^3$. A more elaborate discussion of the nonfactorizable effects will
be presented in a separate publication.

Finally we note that for the moments the situation is different. Contributions
that introduce an IR sensitivity to the charm quark mass in the total rate 
may become regular
for the moments. This becomes evident if we consider moments of the partonic
invariant mass such as $\langle (p^2 - m_c^2)^n \rangle$ which remain regular
as $m_c \to 0$ till higher orders in $\Lam$.

\begin{figure}[!h]
\centerline{
\includegraphics[height=5cm]{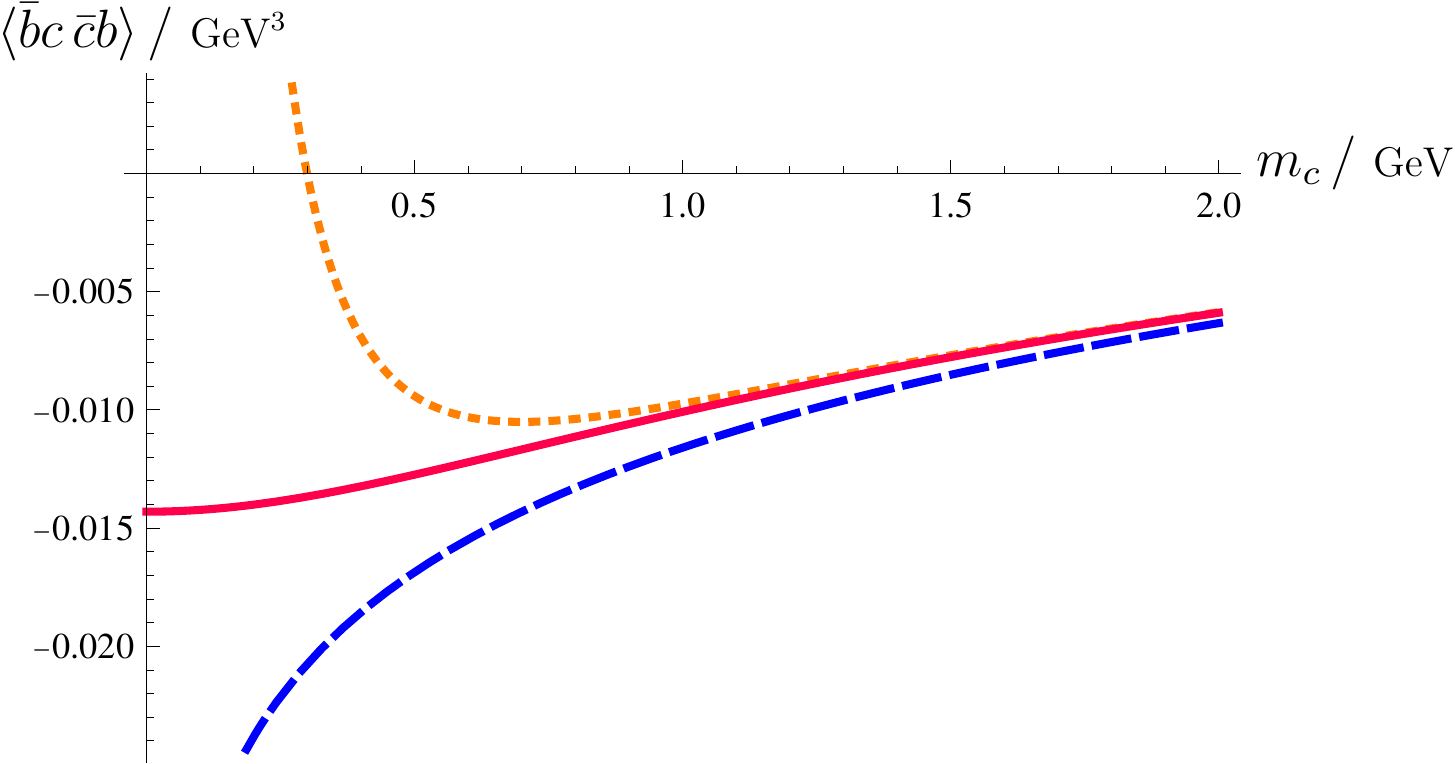}}
\caption{The WA expectation value
$\frac{1}{2M_B}\matel{B}{\bar{b}\gamma^k(1\!-\!\gamma_5)c\, \bar{c}\gamma^k
(1\!-\!\gamma_5)b}{B}_\mu$ as a function of charm mass: the leading contribution
logarithmic in $m_c$ (blue dashed), including additionally the $1/m_c^2$ power
correction (orange dotted), and the complete behavior according to ansatz
Eq.~(\ref{3.263}) (red solid).  We assume the value $\rho_D^3\!=\!0.15\GeV^3$ and
the normalization point $\mu$ is taken $4.6\GeV$.}
\label{WAplot}
\end{figure}

Having at hand the numerical estimates of the higher-dimension expectation
values allows us to refine the model (\ref{3.26}) for the charm-mass
dependence in the IR regime. Since the $1/(m_b^3 m_c^2)$ corrections
calculated in this ansatz are fixed in terms of parameter $\Lambda$, we assign
the latter the value which would reproduce these leading corrections. This
yields
\beq
\Lambda^2\equiv M_*^2= \frac{\tilde f_1}{5\rho_D^3}-\frac{\tilde
f_2}{10\rho_D^3}+\frac{13\tilde f_3}{40\rho_D^3}-\frac{\tilde
f_4}{12\rho_D^3}-\frac{\tilde f_5}{24\rho_D^3}
\simeq (0.7\GeV)^2,
\label{3.262}
\eeq
and the model predicts
\beq
(-\delta^{\alpha\beta}\!+\! 
v^\alpha v^\beta)\,\frac{1}{2M_B}\matel{B}{\bar{b}\gamma_\alpha(1\!-\!\gamma_5)c\, 
\bar{c}\gamma_\beta (1\!-\!\gamma_5)b}{B}_\mu \simeq -\frac{\rho_D^3}{4\pi^2} 
\ln{\frac{\mu^2}{m_c^2\!+\!M_*^2}},
\label{3.263}
\eeq
see Fig.~\ref{WAplot}. In this case the correction to 
the width at $m_c^2 \!\ll\! m_b^2$ takes the form
\beq
\frac{\delta\,\Gamma_{\rm sl}}{\Gamma_{\rm sl}} \simeq -\frac{8\rho_D^3}{m_b^3}
\left(\ln{\frac{m_b^2}{m_c^2\!+\!M_*^2}}-\frac{77}{48}\right)
\label{3.264}
\eeq
where the constant term accounts for the explicit UV contribution in
this limit. This model illustrates to which extent the charm quark may be
considered heavy in this context.

It should be acknowledged that Eqs.~(\ref{3.26}), (\ref{3.262}) and
(\ref{3.264}) are only a reasonable model for the effects at intermediate to
small $m_c$. In fact, the size of the effective mass scale $M_*$ as determined
by matching the leading charm power corrections may not be fully universal: it
depends on the particular Lorentz structure of the weak vertices. It would
likewise differ, say, in weak $B^*$ decays -- although numerically such a
variation in $M_*$ would be insignificant.

\section{Facing up to Weak Annihilation}

Our preceding discussion can suggest a dual description of IR
charm effects that allows insights into a higher-order OPE calculation of the
heavy-to-light case $B \to X_u \ell \bar{\nu}_\ell$. 
More specifically 
the limiting case $m_c \tto 0$ is of relevance when treating the
heavy-to-light case beyond order $1/m_b^2$. The inherent IR divergence of the
standard calculation for $m_c/m_b\tto 0$ underlies the importance of the four
quark operator matrix element in the heavy-to-light case, which is usually
called WA contribution. More precisely, here it corresponds
to its valence-quark    
insensitive piece which affects semileptonic $B^+$ and $B^0$ decays equally. 
We refer to it as non-valence WA.

The important question here is the potential scale of the non-valence WA
contributions, once the effect of the Darwin operator has been separated
out.\footnote{The difference due to the valence part of WA in 
$b\tto u\,\ell\nu$ can be
experimentally probed by analyzing the difference in the semileptonic spectra
spectra of charged and neutral $B$ decays.} A direct computation based on the
method discussed above is not possible. Yet one may try to estimate the
natural scale expected for WA by approaching the massless case from the heavy
quark side. For this purpose we use the model suggested in the previous
sections. In spite of its admitted oversimplification, we can be sure that the
difference between this ansatz and the actual QCD contribution remains finite
at any $m_c$ including the limit $m_c\tto 0$. The model has the advantage of
being sufficiently accurate when extrapolating down from the side of
intermediately heavy charm quarks.

It is therefore plausible that setting $m_c\!=\!0$ and applying this model we
do not stray far away from the leading effect of the non-valence component of
WA in $b\tto u\,\ell \nu$ decays (which is its only contribution
in $B_d$ decays);\footnote{Paper \cite{WA} which first analyzed the effects of
generalized WA in semileptonic decays, focussed on the differences between
mesons with different spectators and therefore explicitly subtracted the
$\bar{b}u \,\bar{u}b$ expectation value in $B_d$ from that in $B^-$. Following
an earlier classification of the preasymptotic power corrections to the
inclusive widths it referred to this as the spectator-dependent correction, a
terminology continued to a number of later publications. The valence and
non-valence effects separately were considered in Ref.~\cite{four} where
valence effects were sometimes also called `spectator' contributions. The
importance of the non-valence WA for light quarks, although conjectured
already back in 1994, was emphasized in Ref.~\cite{Vub}. There the `singlet'
and the WA proper pieces referred to the average and the difference of the
valence and non-valence components.} this guesstimate, though, cannot be
validated in the context of the $1/m_c$ expansion examined here. In
some sense such an assumption implies that no `phase transition' occurs when
going down in mass  from heavy to light quarks. We know such a phase
transition takes place in the QCD vacuum, yet it may not necessarily be
important for the expectation values over $B$ meson states. One may a priori
expect larger WA effects in $b\tto u\,\ell \nu$ coming from its
flavor-specific piece manifesting itself in the decays of charged $B$-mesons.

Using the model of the previous sections to
interpolate between the regimes of heavy and light charm we get an estimate 
\beq
\frac{1}{2M_B}\matel{B_d}{\bar b \vec\gamma (1\!-\!\gamma_5)u \,\bar u
\vec \gamma (1\!-\!\gamma_5) b} {B_d} \simeq  -\frac{\rho_D^3}{4\pi^2}
\ln{\!\left(\frac{\mu^2}{M_*^2}\!+\!1\right)}
\approx -0.005\GeV^3,
\label{310}
\eeq
with $\mu$ denoting the normalization scale. 
This is an educated guess and cannot guarantee to yield even the correct 
sign of the effect at $\mu\lsim 1\GeV$. The negative sign physically
means that the propagation of the soft $u$ quark (projected onto the 
spin state specified by the Lorentz structure in question) is suppressed
compared to free propagation. Taken at face value, the WA expectation value in
Eq.~(\ref{310}) would yield the isoscalar shift in the semileptonic $B$ decay
width 
\beq
\frac{\delta\Gamma^{\rm WA}_{\rm sl}(b\tto u)}
{\Gamma_{\rm sl}(b\tto u)}
 \simeq  -\frac{8\rho_D^3}{m_b^3}
\ln{\!\left(\frac{\mu^2}{M_*^2}\!+\!1\right)}
\approx -0.015\;.
\label{312}
\eeq

A more refined way to assess non-valence WA for $b\tto u\,\ell\nu$ would be to
consider the real massless case for the final-state quark, $m_u\!=\!0$, yet to
introduce an IR cutoff via the kinematic restriction Eq.~(\ref{3.20}). The
first term in Eq.~(\ref{3.1}) representing the `UV' piece of the $\bar{b}u
\,\bar{u}b$ expectation value from the domain of quark momenta above $\mu$ is
then calculated in the direct way of Sect.~2 (with $m_c\tto 0$). The result is
expressed in terms of the same five expectation values, with the coefficients
scaling as $1/\mu^2$; yet they would combine to yield in general a different
combination of the operators and, consequently, a different number.
Evaluating the result with $\mu\!\approx\!0.6\GeV$ would provide an
estimate of the minimal natural scale of non-valence WA.

Even such an estimate would be admittedly incomplete; beyond its lack of
precision in evolving the $\mu$ dependence to as low a value as $0.6\GeV$, the
total WA should also include the $\matel{B}{\bar{b}u
\,\bar{u}b}{B}_{<0.6\,{\rm GeV}}$, the last term in Eq.~(\ref{3.1}). The
contribution from the low momenta plausibly exceeds the numerical estimate
above, and may even change the overall sign. However, it would be unnatural to
allow the contributions from physically distinct domains of low and high
momenta to show significant cancellations. Therefore, we would view the thus
obtained estimate as a firmer lower bound on the scale of non-valence WA in
$b\tto u\, \ell\nu$.

One may anticipate a potentially more significant effect for the `valence'
part of WA which describes this effect in $\Gamma(B^+\tto X_u\,\ell\nu)$. On
physical grounds we expect this contribution to contain a piece independent of
the non-valence WA and not related to something that can be traced from the
$b\tto c\,\ell\nu$ decays in the limit of small charm mass. In the formal
derivation of relating  $c(x)\bar{c}(0)$ in 
Eq.~(\ref{3.1}) to charm Green
functions it would be associated with an additional term. That term appears 
through the
contraction of the quark fields with those of the same flavor whose presence
is required in the interpolating currents to produce the initial and to
annihilate the final $B$ meson state.

On the other hand, as pointed out in Refs.~\cite{WA,Ds} and exploited in later
papers \cite{Vub}, \cite{four}, one may infer certain information about WA,
and in particular about valence WA from the $D$ meson decays. The charm quark
is marginally heavy to apply precision heavy quark expansions to its
decays. Extrapolation from charm to beauty may thus be semi-quantitative at
best, yet it should provide some constraint on the expected scale of WA's
physical implementation. In particular, we want to point out that
the preliminary CLEO-c \cite{CLEOc} data on the semileptonic branching
fraction of $D_s$ indicate the presence of a {\sl destructive}
spectator-related  WA
contribution of around $20\%$; it must be the result of non-factorizable
four-quark expectation values. In interpreting this observation
one has to address the question of $SU(3)$-breaking in the leading
nonperturbative corrections described by the kinetic and chromomagnetic
operators (and, possibly, in the Darwin expectation value).

While semileptonic `valence' WA cannot occur at all in $D^0$ decays, it can
contribute in $D^+$ and $D_s$ decays on the Cabibbo suppressed and Cabibbo
allowed levels, respectively. Assuming possible $SU(3)$ breaking to be under
control in WA proper, we conclude that the observed difference of the total
semileptonic widths for $D_s$ and $D^0$
\beq
\frac{\Gamma_{\rm sl}(D_s)}{\Gamma_{\rm sl}(D^0)}=
\frac{{\rm Br}_{\rm sl}(D_s)}{{\rm Br}_{\rm sl}(D^0)}
\frac{\tau_{D^0}}{\tau_{D_s}}\simeq 0.81
\label{130}
\eeq
must be dominated by valence WA in $D_s$. 
To describe the pattern on the WA effects in $D$ mesons we may adopt the 
nomenclature for the generalized
`annihilation' correction following Ref.~\cite{WA}: 
The {\sl valence} WA for a particular transition
$c\tto q\,\ell\nu$ ($q\!=\!s$ or $d$), ${\rm WA}_q^{\rm val}$ refers to the
{\sf difference} in the matrix elements between $D_q$ and $D^0$.  The {\sl
nonvalence}  ${\rm WA}_d^{\rm n\,val}$ is directly the expectation value in
the (non-strange) state $D^0$ containing no valence $d$ quark. To allow for
$SU(3)$ asymmetry we also have to distinguish ${\rm WA}_d^{\rm n\,val}$ from 
${\rm WA}_d^{\rm n\,val\,(s)}$ where it is considered in the strange $D_s$
state. The WA operators above may be general products like $\bar{c}q\,
\bar{q}c$, either local or nonlocal. (For the decaying
quark being heavy enough, like in $B$-mesons, it would be sufficient to 
include only the leading local four-quark operators.)

With this convention, we in general have in $D$ mesons 
\bea
\nonumber
\Gamma_{\rm sl}(D^+)- \Gamma_{\rm sl}(D^0) &\msp{-4}=\msp{-4}& 
\sin^2{\theta_c}\cdot {\rm
WA}_d^{\rm val}\\
\Gamma_{\rm sl}(D_s)- \Gamma_{\rm sl}(D^0) &\msp{-4}=\msp{-4}&
\cos^2{\theta_c}\cdot 
{\rm WA}_s^{\rm val}-\sin^2{\theta_c}\left[ {\rm WA}_d^{\rm n\,val}- 
{\rm WA}_d^{\rm n\,val\,(s)}
\right] + \Delta_{SU(3)}.
\label{132}
\eea
By introducing the subscript
marking the $d$ or $s$ flavor we have explicitly allowed for the $SU(3)$
breaking in the expectation values due to the different spectator in a meson
or in the light quark field flavor in the corresponding operator. 
We have still neglected the explicit short-distance $SU(3)$-breaking $\propto
\!m_s^2$ in the coefficient functions emerging due to the larger $m_s$ in the
hard quark Green functions; it is expected to be strongly
suppressed. $\Delta_{SU(3)}$ in Eq.~(\ref{132}) therefore denotes only the
shift related to the $SU(3)$ violation in the (flavor-singlet) 
nonperturbative expectation
values between the strange and non-strange heavy meson states. The analysis
suggests that these effects are numerically suppressed for the kinetic and the
chromomagnetic operators and should not exceed $5\%$ level in the widths. Then
the bulk of the difference in Eq.~(\ref{130}) should be equated with the
valence component of WA, at least if $SU(3)$ violation in it is not too
strong.

Translating these relations for WA from charm to beauty is associated with
significant uncertainties due to a potentially poor representation of the
contributions to the inclusive width for charm by the (truncated) OPE; for
$m_c$ is manifestly not large enough for a precision treatment.  Relating WA
for $B$ decays to the expectation values of the $D\!=\!6$ operators can be
done with acceptable theoretical accuracy. Yet expressing the WA contributions
in Eqs.~(\ref{132}) through the analogous local expectation values in $D$
mesons is subjected to large corrections -- first of all from the
corresponding higher-dimension operators with additional derivatives. Related
to this is the short-distance `hybrid' \cite{VShybrid,mirage} renormalization
of the operators in question from the scale of charm to beauty; in this case
it may be even not fully perturbative.

Bearing in mind these potential caveats we nevertheless use this line of
reasoning to estimate the expected size of the valence WA in the semileptonic
$b\tto u$ width of charged $B$ meson:
\beq
\frac{\Gamma(B^+\tto X_{\rm light}\,\ell\nu)-\Gamma(B^0\tto X_{\rm light}\,\ell\nu) }{\Gamma(B\tto
X_{\rm light}\,\ell\nu)} \approx -(0.005 \mbox{ to } 0.01)
\label{136}
\eeq
which is similar in magnitude, yet still below our estimates for the 
minimal scale of the non-valence WA.

\section{Summary and Outlook}
\label{OUT}

The main result of this study is that the OPE for inclusive $B \to X_c \,\ell
\bar{\nu}_\ell$ contains terms with an infrared sensitivity to the charm-quark
mass. Although this has been known, a complete discussion of these
so-called ``intrinsic charm'' contributions had not been presented. 
We have given such 
a discussion here in the context of two theoretical frameworks or `roads' for 
removing $c$ quarks from the dynamical degrees of freedom that a 
priori appear different, yet in the end yield identical results. 

We have shown that starting at $1/m_b^3$ the standard OPE for $B \tto X_c
\,\ell \bar{\nu}_\ell$ exhibits terms of the form 
$1/m_b^m \!\times\! 1/m_c^n$ where at
tree level only even $n$ and odd $m$ appear. The matrix elements of local
operators parametrizing their nonperturbative input always contain
gluon-field-strength operators and their covariant derivatives; in turn the
residual momentum of the $b$ quark does not enter here.

We have performed a detailed analysis of the contributions of the form
$1/m_b^3 \!\times\! 1/m_c^2$ at tree level, which is needed to complete the OPE
calculation of $B \to X_c \ell \bar{\nu}_\ell$ up to order $1/m_b^4$, since
parametrically $1/m_b^3 \!\times\! \Lambda_{\rm QCD} /m_c^2$ is of the same
order. The numerical estimates confirm the results presented in
Ref.~\cite{Bigi:2005bh}. 

The conclusion for $B \tto X_c \,\ell \bar{\nu}_\ell$ is that a calculation to
order $1/m_b^n$ has to include also the terms of order $1/m_b^{n-k} \!\times\!
1/(m_c^2)^k$; furthermore, including radiative corrections one obtains also
contributions of the order $1/m_b^m \times \alpha_s (m_c) / m_c^k$ where $k$
can also be odd. The lowest terms of this kind are of order $1/m_b^3 \times
\alpha_s (m_c) / m_c$ and have been considered in Ref.~\cite{Bigi:2005bh}.

These effects are of considerable theoretical interest with respect to
subtleties that can arise in nonperturbative dynamics, yet they go beyond it
towards more pragmatic goals: they help to validate the goal of reducing the
theoretical uncertainty in extracting $|V_{cb}|$ from $B \tto X_c\,\ell \nu$
to the $1\%$ level; achieving such a goal is of obvious interest for the
theoretical treatment of $B$ decays -- yet also for a
proper interpretation of the ultra-rare decays $K \tto \pi \,\nu \bar \nu$.
Their amplitudes have been calculated with high accuracy in terms
of $m_c$ and $V^*_{ts} V_{td}$ \cite{burasetal}. Their widths thus scale with
$V_{cb}^4$, and the error on the latter is at least a large component in the
stated overall $2\%$ uncertainty.

In $b\tto u\, \ell\nu$ decays the straightforward calculation of the
higher-order power corrections to the total width beyond order $1/m_b^3$ would
yield terms which diverge power-like in the infrared. Yet this does not mean
that one cannot go beyond $1/m_b^3$ order here. The analysis shows that to
calculate $\Gamma_{\rm sl}(b\tto)$ without extra $\alpha_s$-corrections it is
sufficient to introduce the corresponding WA four-quark operators, and then
one should simply discard all the terms formally having inverse powers of
$m_u$.

We have found in the process that analysis of IC effects can inform and focus
our thinking about the possible impact of WA in the heavy-to-light transitions
$B \tto X_u\,\ell \nu $ -- and the extraction of $|V_{ub}|$ there --
and on
its relation to charm decays.  The IC effects for the inclusive distributions
are conceptually similar to (generalized) WA corrections extensively discussed
in connection to the lifetimes of heavy flavors since the late 1970s. Since
the usual beauty hadrons do not contain valence charm quarks, we deal here
with the case of non-valence WA contributions first noted in Ref.~\cite{WA}. A
profound difference with the conventional WA for light quarks is that charm
quarks, 
even soft ones, to the leading approximation can be viewed perturbatively, 
and the nontrivial strong dynamics affect its propagation at the level 
of power corrections
$(\Lam/m_c)^k$, while for light quarks this would not
represent a parametric suppression. Nevertheless, approaching the case of
conventional WA with light flavors from the heavy-mass side and using a model
which naturally interpolates between the regimes of heavy and light quark we
get an estimate
\beq
\frac{\delta \Gamma_{\rm sl}^{{\rm n\,val}}(b\tto u)}{\Gamma_{\rm sl}(b\tto
u)}\approx -0.015\;.
\label{nonvwa}
\eeq
This result should be viewed as an educated guess rather than a real
evaluation; one cannot count even on the 
firm prediction of the sign. It also leaves out the more
intuitive valence WA which, as a matter of fact, historically gave the
phenomenon its name. (Yet, as clarified in Ref.~\cite{mirage}, its
interpretation in the presence of strong interactions is more subtle and may
include interference-type contributions which allow for the net correction to
the width even to become negative.)

We have used the recently reported \cite{CLEOc} measurements of the $D_s$
semileptonic fraction to estimate the significance of 
spectator-related WA in
the KM-suppressed semileptonic width of $B^+$. Taken literally, the correction
turns out to be close to the non-valence case Eq.~(\ref{nonvwa}) or even
somewhat smaller, but still destructive:
\beq
\frac{\delta \Gamma_{\rm sl}^{{\rm val}}(b\tto u)}{\Gamma_{\rm sl}(b\tto
u)}\approx -(0.005 \mbox{~to~} 0.01)\;.
\label{valwa}
\eeq
Since these contributions populate the kinematic domain of small hadronic
invariant mass and energy of the final hadron state, they could have an
amplified impact on the existing determinations of $V_{ub}$ from 
$B \tto X_u \,\ell \bar{\nu}_\ell$.

\subsection*{Acknowledgements}
It is our pleasure to thank A.~Khodjamirian and R.~Zwicky for helpful
discussions and M.~Shifman and M.~Voloshin for bringing Ref.~\cite{CLEOc} 
to our attention. 
This work was supported by the German research foundation DFG under  
contract  MA1187/10-1, by the German Ministry of Research (BMBF),
contracts 05H09PSF and 06SI9192 and by the NSF under the grant number PHY-0807959.


\begin{thebibliography}{99}
\bibitem{Bigi:2005bh}
  I.~I.~Bigi, N.~Uraltsev and R.~Zwicky,
  Eur.\ Phys.\ J.\  C {\bf 50}, 539 (2007)
  [arXiv:hep-ph/0511158].
  %
  %
\bibitem{Breidenbach:2008ua}
  C.~Breidenbach, T.~Feldmann, T.~Mannel and S.~Turczyk,
  Phys.\ Rev.\  D {\bf 78}, 014022 (2008)
  [arXiv:0805.0971 [hep-ph]].
  %
  %
\bibit{brodsky}
S.J.~Brodsky and S.~Gardner,
{\it Phys.\ Rev.}\  {\bf D65} 054016 (2002);\\
S.J.~Brodsky, P.~Hoyer, C.~Peterson and N.~Sakai,
{\it Phys.\ Lett.}\  {\bf B93} 451 (1980);\\
S.J.~Brodsky, C.~Peterson and N.~Sakai,
{\it Phys.\ Rev.}\ {\bf D23} 2745 (1981).

\bibit{WA}
I.I.~Bigi and N.G.~Uraltsev, {\it Nucl.\,Phys.}\ B {\bf423} 
(1994) 33.


\bibitem{Dassinger:2006md}
  B.~M.~Dassinger, T.~Mannel and S.~Turczyk,
  JHEP {\bf 0703}, 087 (2007)
  [arXiv:hep-ph/0611168].

\bibit{SVZFest}
 V.A.~Novikov, M.A.~Shifman, A.I.~Vainshtein and V.I.~Zakharov,
{\it Fortsch.\ Phys.\ }  {\bf 32} (1984) 585.


\bibit{four}
D.\,Pirjol and N.\,Uraltsev, {\it Phys.\,Rev.}\ D {\bf 59} (1999) 034012.


\bibit{Vub}
N.\,Uraltsev, {\it Int.\,Journ.\,Mod.\,Phys.\ }{\bf A14} (1999) 4641.

\bibit{Ds}
I.I.~Bigi and N.G.~Uraltsev,  {\it Zeit.\ f.\ Phys.} C {\bf62} (1994) 623.


\bibit{CLEOc}
\scalebox{.9}{\sf http://indico.cern.ch/materialDisplay.py?contribId=321\&sessionId=38\&materialId=slides\&confId=41044},
slide 34.


\bibit{VShybrid}
M.\,Voloshin and M.\,Shifman, {\em Yad.\,Phys.\ } {\bf 45} (1987) 463 
[{\em Sov.\,J.\,Nucl.\,Phys.\ } {\bf 45} (1987) 292];\\
{\it ZhETF} {\bf 91} (1986) 1180 [{\it JETP} {\bf 64} (1986) 698].


\bibit{mirage}
I.I.~Bigi and N.G.~Uraltsev,
{\it Phys. Lett.}\ B {\bf 280} (1992) 271.


\bibit{burasetal}
U.~Haisch,
PoS {\bf KAON} (2008) 056  [arXiv:0707.3098 [hep-ph]].


\end{thebibliography}
\end{document}